\documentclass[preprint]{elsarticle}
\newcommand\beq{\begin{eqnarray}}
\newcommand\eeq{\end{eqnarray}}
\newcommand\bq{\begin{equation}}
\newcommand\eq{\end{equation}}

\usepackage{amssymb}

\journal{Physic Letters B}

\begin{document}

\begin{frontmatter}


\title{Azimuthal asymmetry of  recoil electrons in  neutrino-electron elastic scattering as signature of neutrino nature}
\author[UWr]{W. Sobk\'ow\corref{cor1}}
 \ead{wieslaw.sobkow@ift.uni.wroc.pl}
\author[UWr]{A. B{\l}aut}
 \ead{arkadiusz.blaut@ift.uni.wroc.pl}
 \cortext[cor1]{Corresponding author}
\address[UWr]{Institute of Theoretical Physics, University of Wroc\l{}aw, Pl. M. Born 9,
\\PL-50-204~Wroc\l{}aw, Poland}



\begin{abstract}
In this paper, we show how a presence of the exotic scalar, tensor weak interactions in addition to the standard vector-axial (V-A) one may help to distinguish the Dirac from Majorana neutrinos in the elastic scattering of (anti)neutrino beam off the unpolarized electrons in the limit of vanishing (anti)neutrino mass.  We assume that the incoming (anti)neutrino beam comes from the polarized muon decay at rest  and is the left-right chiral mixture with  assigned direction of the transversal  spin polarization with respect  to the production plane. We display that the azimuthal asymmetry in the angular distribution of  recoil electrons is generated by the interference terms between the standard and exotic couplings, which are proportional to the transversal  (anti)neutrino  spin polarization and independent of  the neutrino mass. This asymmetry  for the Majorana neutrinos  is distinct from the one for the Dirac neutrinos through the absence of   interference  between the standard   and tensor couplings. Additionally, the interference term between the standard and scalar coupling of the only left chiral neutrinos, absent in the Dirac case, appears.  We also indicate the possibility of utilizing the azimuthal asymmetry measurements to search for the new CP-violating phases. 
Our analysis is model-independent and consistent with the current upper limits on the non-standard couplings.  
\end{abstract}


\end{frontmatter}


\section{Introduction} 
\label{sec1}
One of the fundamental problems in  the neutrino physics is whether the  neutrinos $(\nu)$'s are  the Dirac or Majorana fermions. The question of $\nu$ nature can be probed in the context of non-vanishing $\nu$ mass and of standard vector-axial $(V-A)$ weak interaction with only the  left chiral (LCh) $\nu$'s \cite{SM}, using purely leptonic processes such as the polarized muon decay at rest (PMDaR) or the  neutrino-electron elastic scattering (NEES). There is an alternative way within the massless $\nu$ limit, when one admits the existence of exotic scalar (S), tensor (T), pseudoscalar (P) and $V+A$ weak interactions  of the right chiral (RCh) $\nu$'s (right-handed helicity when $m_{\nu} \rightarrow 0$) in addition to the $V-A$ interaction of the LCh ones in the above processes.  The appropriate tests involving the mass dependence have been proposed by Kayser \cite{Kayser} and Langacker  \cite{Langacker}. It is also worthwhile remarking the others interesting ideas regarding the $\nu$ nature problem 
\cite{Semikoz, Pastor, Barranco, Singh, Gutierrez}. 
One ought to emphasize that at present the neutrinoless double beta decay (NDBD) seems to be the best tool to investigate the $\nu$ nature  
\cite{Majorana}, however the mentioned above  processes may also shed some light on this problem. \\
First tests concerning   the problem of distinguishing between the Dirac and Majorana  $\nu$'s in the limit of vanishing $\nu$ mass, when one departs from the V-A interaction and one allows for the exotic S, T, P  weak interactions in the NEES, have been reported by Rosen \cite{Rosen} and Dass \cite{Dass}. The  leptonic processes are also  suitable to probe  the time reversal violation (TRV) effects. 
It is relevant to   point out that  the existing data  still leaves a small space for the exotic couplings of the interacting RCh $\nu$'s. It is noteworthy  that the effects coming from  the interacting $\nu$'s with right-handed chirality  are also important for interpreting of results on the NDBD \cite{ndbd}. 
Unfortunately, the proposed quantities in \cite{Rosen, Dass} are composed of   the squares of exotic couplings of the RCh  $\nu$'s and  at most of the interferences within exotic couplings, that  are both very tiny.  Furthermore, both  transverse components of electron (positron) spin  polarization and neutrino energy spectrum in the PMDaR   contain only the interference terms between the standard V and non-standard S, T couplings of LCh $\nu$'s. All the eventual interferences  between the standard couplings of LCh  and exotic couplings of RCh  $\nu$'s  vanish, because are proportional to a tiny $\nu$ mass and do not produce the effect. As the current experiments do not detect the RCh $\nu$'s, it seems  meaningful to search for new tools including the linear terms from the exotic couplings that are  independent of the $\nu$ mass, and  obtained in model-independent way. It would enable to compare the predictions of various non-standard schemes   with the experimental data, and  look for the TRV effects. 
The suitable observables could be  the $\nu$ quantities  carrying  information on the transversal components of (anti)neutrino spin polarization, both T-odd and T-even. Presently,  such tests are still not available, because they require the observation of final $\nu$'s, the strong $\nu$ beam coming from the polarized source and  the efficient $\nu$ polarimeters. However, it is worthy of indicating the potential possibilities of experiments of the $\nu$ polarimetry in the connection with the  $\nu$ nature problem, the existence of interacting RCh  $\nu$'s and the non-standard TRV phases predicted by  many extensions of   the SM. Let us recall that the SM can not be viewed as a ultimate theory, because it does not clarify  the origin of  parity violation  at current energies,  the observed baryon asymmetry of universe \cite{barion} through  a single CP-violating phase of the Cabibbo-Kobayashi-Maskawa quark-mixing matrix (CKM) \cite{Kobayashi}, the large hierarchy fermion masses, and  others  fundamental  aspects. This situation  led to the appearance  of various non-standard gauge models including the Majorana $\nu$'s, exotic TRV interactions, mechanisms explaining the origin of fermion generations, masses, mixing and smallness of $\nu$ mass. \\
In this paper, we focus on the elastic  scattering of electron  $\nu$'s ($\overline{\nu}$'s) off the unpolarized electron target and  show in model-independent way how the participation of the exotic S, T couplings of RCh $\nu$'s ($\overline{\nu}$'s) in addition to the standard couplings of LCh ones can be utilized to distinguish the Dirac from Majorana $\nu$'s, and to test the TRV in the limit of vanishing $\nu$ mass. 

\section{Elastic scattering of Dirac electron antineutrinos off unpolarized electrons }
\label{sec2}
We  assume that the incoming Dirac $\overline{\nu}_{e}$ beam comes from the decay of polarized negative muons at rest $(\mu^{-} \rightarrow e^- +
\overline{\nu}_{e} + \nu_{\mu})$ and  is a mixture of left-right chiral states with a fixed direction of the transversal spin polarization with  respect to the production plane. LCh $\overline{\nu}_{e}$'s are  mainly detected by the standard $V-A$ interaction and RCh ones are detected only by the exotic scalar, tensor interactions in the elastic scattering on the unpolarized electron target; $(\overline{\nu}_{e} + e^{-}\rightarrow \overline{\nu}_{e} +e^{-})$. Our scenario admits also the detection of $\overline{\nu}_{e}$'s with left-handed chirality by the non-standard S and T interactions.  The production plane for the $\overline{\nu}_{e}$ beam, shown in Fig. 1,  is spanned by the unit vector 
$\mbox{\boldmath $\hat{\eta}_{\mu}$}$ of the muon polarization and  the $\overline{\nu}_{e}$ 
LAB momentum unit vector ${\bf \hat{q}}$.  In 
this plane, the  vector $\mbox{\boldmath $\hat{\eta}_{\mu}$}$  can 
be expressed, with respect to  $\hat{\bf q}$, as a sum of 
$(\mbox{\boldmath$\hat{\eta}_{\mu}$}\cdot\hat{\bf q}){\bf \hat{q}} $
 and $ \mbox{\boldmath $\eta_{\mu}^{\perp}$} =
\mbox{\boldmath$\hat{\eta}_{\mu}$}-
(\mbox{\boldmath$\hat{\eta}_{\mu}$}\cdot\hat{\bf q}) {\bf \hat{q}}$. 
The reaction (detection) plane is spanned by   the direction of the outgoing electron
momentum  $ \hat{\bf p}_{e}$  and $\hat{\bf q}$, Fig. 1. 
\begin{figure}

\begin{center}
\includegraphics[scale=.5]{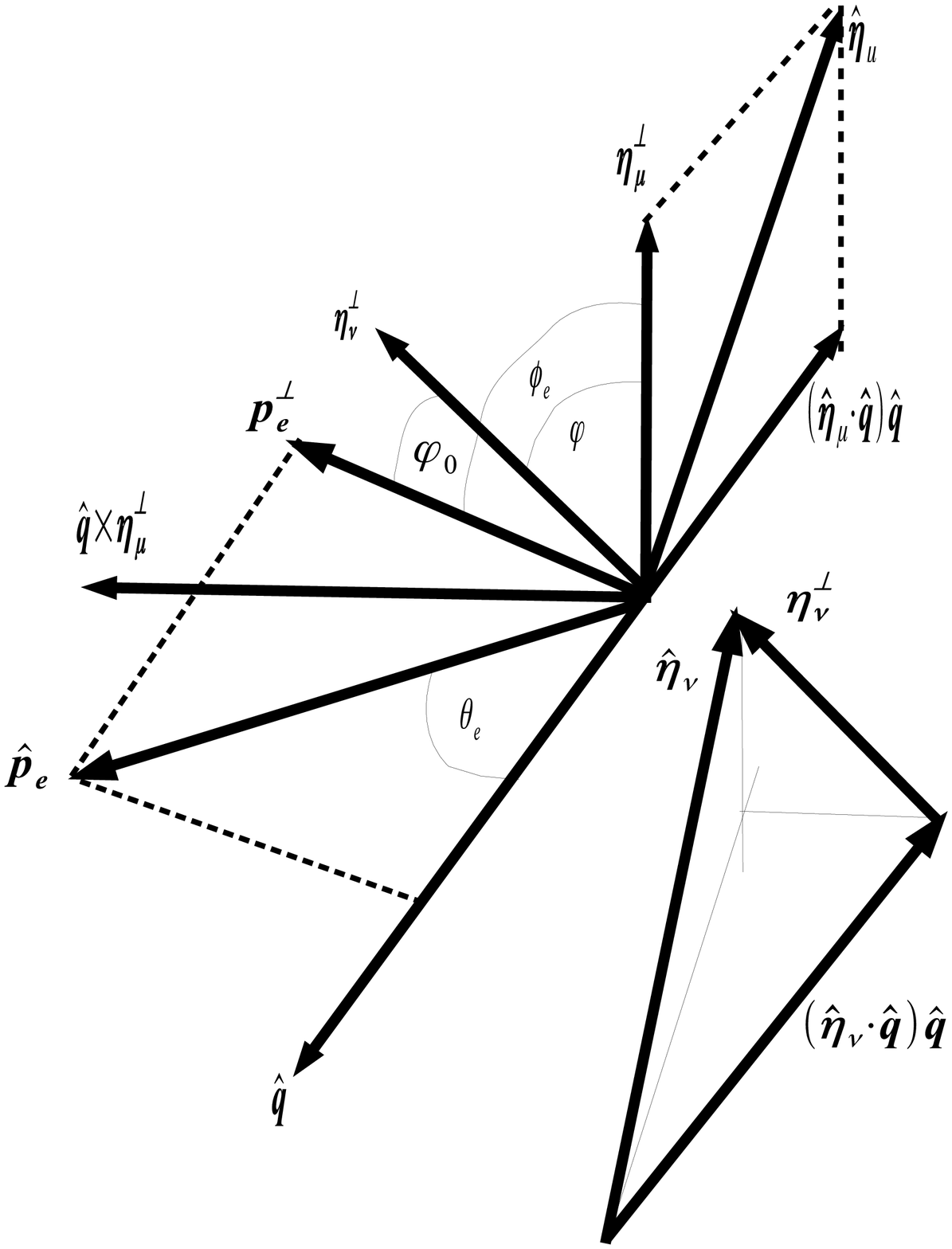}
\end{center}
\caption{Production plane of the
$\overline{\nu}_{e}$ beam is   spanned by the vectors $\mbox{\boldmath $\hat{\eta}_{\mu}$}$  and   ${\bf \hat{q}}$ for  $\mu^{-} \rightarrow e^-
+ \overline{\nu}_{e} + \nu_{\mu}$. Reaction  plane is spanned by  the vectors $\hat{\bf p}_{e}$  and  $\hat{\bf q}$   for $\overline{\nu}_{e} + e^{-}\rightarrow \overline{\nu}_{e} +e^{-}$.    
$\mbox{\boldmath $\hat{\eta}_{\overline{\nu}}$}$ is expressed, with respect to  $\hat{\bf q}$, as a sum of $(\mbox{\boldmath$\hat{\eta}_{\overline{\nu}}$}\cdot\hat{\bf q}){\bf\hat{q}}$ 
and $\mbox{\boldmath $\eta_{\overline{\nu}}^\perp $}$. 
 \label{DMnu}}
\end{figure}
The amplitude for the   $\overline{\nu}_{e}  e^{-}$ scattering   at low energies is as follows:
\beq \label{ampD} M^{D}_{\overline{\nu}_{e} e^{-}}
&=&
-\frac{G_{F}}{\sqrt{2}}\{(\overline{u}_{e'}\gamma^{\alpha}(c_{V}^{L}
- c_{A}^{L}\gamma_{5})u_{e}) (\overline{v}_{\nu_{e}}
\gamma_{\alpha}(1 - \gamma_{5})v_{\nu_{e'}})\\ 
&  & \mbox{} +
c_{S}^{R}(\overline{u}_{e'}u_{e})(\overline{v}_{\nu_{e}}
(1 - \gamma_{5})v_{\nu_{e'}}) \nonumber\\
&& \mbox{} +
\frac{1}{2}c_{T}^{R}(\overline{u}_{e'}\sigma^{\alpha \beta}u_{e})(\overline{v}_{\nu_{e}}
\sigma_{\alpha \beta}(1 - \gamma_{5})v_{\nu_{e'}})\nonumber\\
&&\mbox{} +
c_{S}^{L}(\overline{u}_{e'}u_{e})(\overline{v}_{\nu_{e}}
(1 + \gamma_{5})v_{\nu_{e'}}) \nonumber\\
&& \mbox{} +
\frac{1}{2}c_{T}^{L}(\overline{u}_{e'}\sigma^{\alpha \beta}u_{e})(\overline{v}_{\nu_{e}}
\sigma_{\alpha \beta}(1 + \gamma_{5})v_{\nu_{e'}})
\}\nonumber,  
 \eeq
 where $G_{F} = 1.1663788(7)\times
10^{-5}\,\mbox{GeV}^{-2} (0.6 \; ppm)$ \cite{Mulan} is the Fermi constant. 
  The coupling constants are
denoted with the superscripts $L $ and $R $ as $c_{V}^{L} $, 
$c_{A}^{L}$, $c_{S}^{R, L}$, $c_{T}^{R, L}$ respectively to the incoming $\overline{\nu}_{e}$ 
of left- and right-handed chirality.  Because we take into account  the TRV, all the coupling constants  are complex. 
Calculations are carried out with  use of the covariant 
density matrix   for the polarized initial $\overline{\nu}_{e}$. 
The formula for the projector $\Lambda_{\overline{\nu}}^{(s)}$ in the massless $\overline{\nu}_{e}$ limit   is given by:
\beq
\lim_{m_{\overline{\nu}}\rightarrow 0}\Lambda_{\overline{\nu}}^{(s)} 
 & = & \mbox{} \frac{1}{2}\bigg\{(q^{\mu}\gamma_{\mu})
 \left[1 - \gamma_{5}(\mbox{\boldmath
$\hat{\eta}_{\overline{\nu}}$}\cdot{\bf \hat{q}}) - \gamma_{5}
S^{\prime\perp}\cdot \gamma \right]\bigg\},
\eeq
where $\mbox{\boldmath $\hat{\eta}_{\overline{\nu}}$}$ is the unit 3-vector of
 $\overline{\nu}_{e}$ spin  polarization in its rest frame; $(\mbox{\boldmath$\hat{\eta}_{\overline{\nu}}$}\cdot\hat{\bf q}){\bf\hat{q}}$ is the longitudinal component of $\overline{\nu}_{e}$ spin polarization; 
$\mbox{\boldmath$\eta_{\overline{\nu}}^{\perp}$} = \mbox{\boldmath$\hat{\eta}_{\overline{\nu}}$} - (\mbox{\boldmath
$\hat{\eta}_{\overline{\nu}}$}\cdot{\bf \hat{q}}){\bf \hat{q}} $ is the transversal component of $\overline{\nu}_{e}$  spin polarization; $S^{\prime\perp} =
(0, \mbox{\boldmath $\eta_{\overline{\nu}}^{\perp}$})$.\\ We see
that in spite of the singularities $m_{\overline{\nu}}^{-1}$ in  the Lorentz boosted spin polarization 4-vector of  massive $\overline{\nu}_{e}$ 
$S^\prime$ (in the laboratory frame), the projector $\Lambda_{\overline{\nu}}^{(s)}$  including $\mbox{\boldmath $\eta_{\overline{\nu}}^{\perp}$}$ remains finite \cite{Michel}. One should notice that the last term in $\Lambda_{\overline{\nu}}^{(s)}$ has different $\gamma$-matrix structure from that of the longitudinal polarization contribution. This term will generate  the non-vanishing interferences between the standard  and exotic couplings  in the differential cross section for the  $\overline{\nu}_{e} e^{-}$ scattering. \\
Using the current data  for the muon decay at rest \cite{Data}, we calculate the upper limit on the
magnitude of $\mbox{\boldmath$\eta_{\overline{\nu}}^{\perp}$} $ and lower bound for 
$(\mbox{\boldmath$\hat{\eta}_{\overline{\nu}}$}\cdot\hat{\bf q})$, \cite{Fetscher}:
\beq \label{trlo}
 |\mbox{\boldmath $\eta_{\overline{\nu}}^{\perp}$}|
&=& 2\sqrt{Q_{L}^{\nu}(1-Q_{L}^{\nu})}  \leq 0.537, \\
|\mbox{\boldmath $\hat{\eta}_{\overline{\nu}}$}\cdot\hat{\bf q}| &=& |2
Q_{L}^{\overline{\nu}}-1|  \geq 0.843,
\nonumber\\
Q_{L}^{\overline{\nu}} &=& 1 - \frac{1}{4}(|g_{LR}^S|^{2} + |g_{LL}^S|^{2})-  3|g_{LR}^T|^{2}\geq
0.922.\nonumber
\eeq

\subsection{Azimuthal distribution of recoil electrons in case of Dirac electron antineutrinos}
\label{subsec2.1}
The differential cross section for the scattering of Dirac $\overline{\nu}_{e}$'s off the unpolarized electrons, when  $m_{\overline{\nu}_{e}}\rightarrow 0$, takes the form: 
\beq \label{przekranue} \frac{d^{2}
\sigma}{d y_{e} d \phi_{e}} &=& \bigg( \frac{d^{2} \sigma}{d y_{e} d
\phi_{e}}\bigg)_{(V- A)} +  \bigg( \frac{d^{2} \sigma}{d y_{e} d \phi_{e}}\bigg)_{(S, T)} \\ &&
\mbox{} + \bigg( \frac{d^{2} \sigma}{d y_{e} d \phi_{e}} \bigg)_{(V S)} + \bigg( \frac{d^{2}
\sigma}{d y_{e} d \phi_{e}} \bigg)_{(A T)}, \nonumber  
\\ \label{VA} \bigg( \frac{d^{2} \sigma}{d
y_{e} d \phi_{e}} \bigg)_{(V- A)} &=& B \bigg\{ (1+\mbox{\boldmath
$\hat{\eta}_{\overline{\nu}}$}\cdot\hat{\bf q}
) \bigg[|c_{V}^{L} - c_{A}^{L}|^{2} + |c_{V}^{L}+ c_{A}^{L}|^{2}(1-y_{e})^{2}\\
&& \mbox{} - \frac{m_{e}y_{e}}{E_{\nu}}\left(|c_{V}^{L}|^{2} - |c_{A}^{L}|^{2}\right) \bigg] \bigg\},
\nonumber \\ 
\bigg(\frac{d^{2} \sigma}{d y_{e} d \phi_{e}}\bigg)_{(S, T)} &=& \mbox{}
B\Bigg\{ (1-\mbox{\boldmath $\hat{\eta}_{\overline{\nu}}$}\cdot\hat{\bf
q})\bigg[ \frac{1}{2}y_{e}\left(y_{e}+2\frac{m_{e}}{E_{\nu}}\right)
  |c_{S}^{R}|^{2}  \\
&& \mbox{} + \left((2-y_{e})^2 -\frac{m_{e}}{E_{\nu}}y_{e}\right)|{c_{T}^{R}}|^{2}
- y_{e}(y_{e}-2)Re(c_{S}^{R}c_{T}^{*R}) \bigg] \nonumber\\
&& \mbox{} +  (1+\mbox{\boldmath $\hat{\eta}_{\overline{\nu}}$}\cdot\hat{\bf
q})\bigg[ \frac{1}{2}y_{e}\left(y_{e}+2\frac{m_{e}}{E_{\nu}}\right) |c_{S}^{L}|^{2}  \nonumber\\
&& \mbox{} + \left((2-y_{e})^2 -\frac{m_{e}}{E_{\nu}}y_{e}\right)|{c_{T}^{L}}|^{2}
- y_{e}(y_{e}-2)Re(c_{S}^{L}c_{T}^{*L}) \bigg] \Bigg\}, \nonumber\\
 \label{VLSR} \bigg(\frac{d^{2}\sigma}{d y_{e} d \phi_{e}}\bigg)_{(V S)} &=& \mbox{}
B \bigg\{ - 4 \sqrt{y_{e}(y_{e}+2\frac{m_{e}}{E_{\nu}})}\bigg[\mbox{\boldmath
$\eta_{\overline{\nu}}^{ \perp}$}\cdot({\bf \hat{p}_{e} \times
\hat{q}})Im(c_{V}^{L}c_{S}^{R*}) \\ && \mbox{} + (\mbox{\boldmath
$\eta_{\overline{\nu}}^{ \perp}$}\cdot {\bf \hat{p}_{e}}) Re(c_{V}^{L}c_{S}^{R*})\bigg]
\bigg\},\nonumber \\
\label{ALTR} \bigg( \frac{d^{2} \sigma}{d y_{e} d \phi_{e}}\bigg)_{(A T)} &=& B \bigg\{ 
- 2 \sqrt{y_{e}(y_{e}+2\frac{m_{e}}{E_{\nu}})}\bigg[\mbox{\boldmath
$\eta_{\overline{\nu}}^{ \perp}$}\cdot({\bf \hat{p}_{e} \times
\hat{q}})Im(c_{A}^{L}c_{T}^{R*}) \\ && \mbox{} + (\mbox{\boldmath
$\eta_{\overline{\nu}}^{ \perp}$}\cdot {\bf \hat{p}_{e}}) Re(c_{A}^{L}c_{T}^{R*})\bigg]
\bigg\}.\nonumber \\
  y_{e} & \equiv &
\frac{T_{e}}{E_{\nu}}=\frac{m_{e}}{E_{\nu}}\frac{2 cos^{2}\theta_{e}}
{(1+\frac{m_{e}}{E_{\nu}})^{2}-cos^{2}\theta_{e}} \eeq is the ratio of the
kinetic energy of the recoil electron $T_{e}$  to the incoming antineutrino 
energy $E_{\nu}$; $B\equiv \left(E_{\nu} m_{e}/4\pi^2\right) \left(G_{F}^{2}/2\right)$;
$\theta_{e}$ is the angle between 
 $ \hat{\bf p}_{e}$  and $\hat{\bf q}$ (recoil electron scattering angle); $m_{e}$ is the
electron mass; 
$\phi_{e}$ is the 
angle between the production plane and the reaction plane (azimuthal angle  of
outgoing electron momentum).
 We see that the interference terms, Eqs. (\ref{VLSR}, \ref{ALTR}), between the
standard $c_{V, A}^{L}$ and exotic $c_{S, T}^{R}$ couplings  survive in the massless
$\overline{\nu}_{e}$ limit. There are no interferences between $c_{V, A}^{L}$ and $c_{S, T}^{L}$ couplings of the LCh $\overline{\nu}_{e}$'s for $m_{\overline{\nu}}\rightarrow 0$.  It can be noticed that the  interferences include only the contributions from
the transverse components of the  $\overline{\nu}_{e}$ spin polarization, both $T$-even
and $T$-odd: 
\beq \label{inter} (\frac{d^{2} \sigma}{d y_{e} d
\phi_{e}})_{(VS)} + (\frac{d^{2} \sigma}{d y_{e} d \phi_{e}})_{(AT)} &=& - B
|\mbox{\boldmath $\eta_{\overline{\nu}}^{ \perp}$}|
\sqrt{\frac{m_{e}}{E_{\nu}}y_{e}[2-(2+\frac{m_{e}}{E_{\nu}})y_{e}]}\\
\cdot \bigg\{ 4 |c_{V}^{L}||c_{S}^{R}|cos(\phi+\beta_{SV}-\phi_e) & + &
2|c_{A}^{L}||c_{T}^{R}|cos(\phi+\beta_{TA}-\phi_e) \bigg\}, \nonumber \eeq 
where 
$\phi$ is the angle between  $\mbox{\boldmath
$\eta_{\overline{\nu}}^{\perp}$}$ and $\mbox{\boldmath$\eta_{\mu}^{\perp}$}$ only; $\phi_0 = \phi- \phi_e$ is the angle between ${\bf p}_{e}^{\perp}$ and $\mbox{\boldmath
$\eta_{\overline{\nu}}^{\perp}$}$, Fig.1; 
$\beta_{VS} \equiv \beta_{V}^{L} - \beta_{S}^{R}, \beta_{AT}
\equiv \beta_{A}^{L} - \beta_{T}^{R} $ are the relative phases between the
$c_{V}^{L}, c_{S}^{R}$ and $ c_{A}^{L}, c_{T}^{R}$ couplings, respectively. 
The relative phases $\beta_{VS}, \beta_{AT}$  
different from $0, \pi$ would indicate the CPV
 in the NEES.  We see that even when $m_{\overline{\nu}}\rightarrow 0$ the helicity structure of interaction vertices may allow for a helicity flip provided the quantity $\mbox{\boldmath $\eta_{\overline{\nu}}^{ \perp}$}$, which is left invariant under Lorentz boost. 
For the standard V-A interaction, Eq. (\ref{VA}), there is no dependence on the $\phi_{e}$, it means that the azimuthal distribution is symmetric. 
In the case of the mixture of  LCh and RCh $\overline{\nu}_{e}$'s,  the interference terms are  proportional to $|\mbox{\boldmath$\eta_{\nu}^{\perp}$}|$ and  depend on  $\phi_e$. It generates the azimuthal asymmetry in the angular distribution of scattered electrons.   
\\ Using the experimental values of  standard couplings: $c_{V}^{L}= 1+ (-0.04 \pm 0.015), c_{A}^{L}= 1+ (-0.507 \pm 0.014)$   
 \cite{Data}, we find the upper limits on the exotic couplings in the NEES: $|c_{S}^{L}|\leq 0.501, |c_{S}^{R}|\leq 0.486, |c_{T}^{L}|\leq 0.003, |c_{T}^{R}|\leq 0.489$, simultaneously shifting the standard values to new ones $c_{V'}^{L}=0.945, c_{A'}^{L}=0.479$, which still lie within the experimental bars. 
  If one uses the above limits, one gets the same magnitude of total cross section as in the standard case.   Now, we calculate the upper limits on the azimuthal asymmetry between the $(0,\pi)$ and $(\pi, 2\pi)$ angles (up-down asymmetry) using the upper limit on  $|\mbox{\boldmath$\eta_{\overline{\nu}}^{\perp}$}| $ and lower bound on  
$(\mbox{\boldmath$\hat{\eta}_{\overline{\nu}}$}\cdot\hat{\bf q})$: 
\beq
A(\phi_{e'}) &=& \frac{\int_{\phi_{e'}}^{\phi_{e'} + \pi} \frac{d\sigma}{d\phi_{e}} d \phi_{e} - \int_{\phi_{e'} + \pi}^{\phi_{e'} + 2\pi} \frac{d\sigma}{d\phi_{e}} d \phi_{e}}{\int_{\phi_{e'}}^{\phi_{e'} + \pi} \frac{d\sigma}{d\phi_{e}} d \phi_{e} + \int_{\phi_{e'} + \pi}^{\phi_{e'} + 2\pi} \frac{d\sigma}{d\phi_{e}} d \phi_{e}}.
\eeq
 The  differential cross section is integrated over $y_e \in [0, 1/(1+(m_e/2E_\nu))]$. 
We get for the case of TRV and time reversal conservation (TRC) with  $E_{\overline{\nu}}=50 \, MeV$, respectively: 
\beq
A_{D}(\phi_{e'})^{T-viol} &=& 0.024 \,cos(\phi- \phi_{e'}) \,\,\mbox{$for$} \,\,\beta_{VS}=\frac{\pi}{2}, \beta_{AT}=\frac{\pi}{2}, \\
A_{D}(\phi_{e'})^{T-cons} &=&  -0.024 \,sin(\phi-\phi_{e'})\,\, \mbox{$for$} \,\, \beta_{VS}=0, \beta_{AT}=0.
\eeq


\section{Elastic scattering of Majorana electron neutrinos off unpolarized electrons} 
\label{sec3}
The amplitude for the elastic  scattering of  the Majorana electron neutrinos ($\nu_{e}$'s) on the unpolarized electrons  at low energies has the form:
\beq \label{ampM} M^{M}_{\nu_{e} e^{-}}
 & = & \mbox{} \frac{G_{F}}{\sqrt{2}}\{(- 2)(\overline{u}_{e'}\gamma^{\alpha}(c_{V}
- c_{A}\gamma_{5})u_{e}) (\overline{u}_{\nu_{e'}}\gamma_{\alpha}\gamma_{5} u_{\nu_{e}})  \\ 
&  &
\mbox{} + 2 c_{S}^{R}(\overline{u}_{e'}u_{e})(\overline{u}_{\nu_{e'}}
(1 + \gamma_{5})u_{\nu_{e}}) \nonumber \\
&& \mbox{} + 2 c_{S}^{L}(\overline{u}_{e'}u_{e})(\overline{u}_{\nu_{e'}}
(1 - \gamma_{5})u_{\nu_{e}})
\}.  \nonumber 
 \eeq
 We see that the above matrix element in the neutrino part does not contain the contribution from the  V  and T  interactions in contrast to the Dirac case, where both terms partake.  In addition, the  A and S  contributions are multiplied by factor 2. This arises from the fact that the Majorana neutrino is described by the self-conjugate field. Absence of the index L for $c_V, c_A$ couplings means that both LCh and RCh $\nu_{e}$'s may participate in the standard A interaction of Majorana $\nu_{e}$'s. In consequence, the new term with the interference between $c_V$ and $c_{S}^{L}$ couplings in the differential cross section appears. Such interference vanishes in the Dirac case.   All the couplings are assumed to be complex as for the Dirac case.  The others assumptions concerning the production of $\nu_{e}$ beam and the detection of $\nu_{e}$'s by the interaction with the unpolarized electron target are the same as in the Dirac scenario.  
 \subsection{Azimuthal distribution of recoil electrons in case of Majorana neutrinos}
 \label{sec3.1}
The azimuthal distribution of the recoil electrons for the scattering of Majorana  $ \nu_{e}$'s on the unpolarized electrons with  $m_{\overline{\nu}}\rightarrow 0$     has the form: 

 \beq \label{przekMaj} \frac{d^{2} \sigma}{d y_{e} d \phi_{e}} &=& \bigg( \frac{d^{2} \sigma}{d y_{e} d
\phi_{e}}\bigg)_{(V- A)} + 
+ \bigg( \frac{d^{2} \sigma}{d y_{e} d \phi_{e}}\bigg)_{(S_{L}, S_{R})} \\ 
&& \mbox{} + \bigg( \frac{d^{2} \sigma}{d y_{e} d \phi_{e}} \bigg)_{(V S_{R})} +  \bigg( \frac{d^{2} \sigma}{d y_{e} d \phi_{e}} \bigg)_{(V S_{L})},  \nonumber \\ 
\label{MAL} \bigg( \frac{d^{2} \sigma}{d y_{e} d \phi_{e}} \bigg)_{(V- A)} &=& \mbox{}  B \bigg\{ |c_{V} - c_{A}|^{2} (2+ (1-\mbox{\boldmath
$\hat{\eta}_{\nu}$}\cdot\hat{\bf q})(y_{e}-2)y_{e})\\
&& \mbox{} + |c_{V} + c_{A}|^{2} (2+ (1+\mbox{\boldmath
$\hat{\eta}_{\nu}$}\cdot\hat{\bf q})(y_{e}-2)y_{e})\nonumber\\
&& \mbox{}- \frac{2 m_{e}y_{e}}{E_{\nu}}\left(|c_{V}|^{2} - |c_{A}|^{2}\right)\bigg\},\nonumber\\ 
\label{MSLR}\bigg(\frac{d^{2} \sigma}{d y_{e} d \phi_{e}}\bigg)_{(S_L,S_R)} &=& \mbox{}
B\bigg\{  2 y_{e}\left(y_{e}+2\frac{m_{e}}{E_{\nu}}\right)
\bigg[ (1+\mbox{\boldmath $\hat{\eta}_{\nu}$}\cdot\hat{\bf q})|c_{S}^{R}|^{2} \\
&& \mbox{} + (1-\mbox{\boldmath $\hat{\eta}_{\nu}$}\cdot\hat{\bf q})|c_{S}^{L}|^{2}\bigg]\bigg\},\nonumber\\   
 \label{MVLSR} \bigg(\frac{d^{2}\sigma}{d y_{e} d \phi_{e}}\bigg)_{(V S_R)} &=& \mbox{} B \bigg\{ 8 \sqrt{y_{e}(y_{e}+2\frac{m_{e}}{E_{\nu}})}\bigg[- \mbox{\boldmath
$\eta_{\nu}^{\perp}$}\cdot({\bf \hat{p}_{e} \times
\hat{q}})Im(c_{V}c_{S}^{R*}) \nonumber \\ && \mbox{} + (\mbox{\boldmath
$\eta_{\nu}^{ \perp}$}\cdot {\bf \hat{p}_{e}}) Re(c_{V}c_{S}^{R*})\bigg] 
\bigg\}, \\
 \label{MVLSL} \bigg(\noindent\frac{d^{2}\sigma}{d y_{e} d \phi_{e}}\bigg)_{(V S_L)} &=& \mbox{}B \bigg\{ 8 \sqrt{y_{e}(y_{e}+2\frac{m_{e}}{E_{\nu}})}\bigg[ \mbox{\boldmath
$\eta_{\nu}^{ \perp}$}\cdot({\bf \hat{p}_{e} \times
\hat{q}})Im(c_{V}c_{S}^{L*}) \\ && \mbox{} + (\mbox{\boldmath
$\eta_{\nu}^{ \perp}$}\cdot {\bf \hat{p}_{e}}) Re(c_{V}c_{S}^{L*})\bigg]
\bigg\}. \nonumber
\eeq
The interference contribution can be written down as follows: 
\beq \label{Majinter} (\frac{d^{2} \sigma}{d y_{e} d
\phi_{e}})_{(V S_R)} + (\frac{d^{2} \sigma}{d y_{e} d \phi_{e}})_{(V S_L)} &=& 8 B
|\mbox{\boldmath $\eta_{\nu}^{ \perp}$}|
\sqrt{\frac{m_{e}}{E_{\nu}}y_{e}[2-(2+\frac{m_{e}}{E_{\nu}})y_{e}]}\\
\cdot |c_{V}|\bigg\{  |c_{S}^{R}|cos(\phi - \alpha_{VS_R}-\phi_e) & + &
|c_{S}^{L}|cos(\phi+\alpha_{VS_L}-\phi_e) \bigg\}, \nonumber \eeq 
where $\alpha_{VS_R} \equiv \alpha_{V} - \alpha_{S}^{R}, \alpha_{VS_L}
\equiv \alpha_{V} - \alpha_{S}^{L} $ are the relative phases between the
$c_{V}, c_{S}^{R}$ and $ c_{V}, c_{S}^{L}$ couplings, respectively. 
Now, we  calculate the upper limits on the azimuthal asymmetry between  $(0,\pi)$ and $(\pi, 2\pi)$ angles for the TRV and TRC, using the same limits as for the Dirac case and $E_{\nu}= 50 \, MeV$: 
\beq
 A_{M}(\phi_{e'}^{T-viol}) &\leq & -0.062\, cos(\phi-\phi_{e'}) \,\,\mbox{$for$} \,\,\alpha_{VS_R}=\frac{\pi}{2}, \alpha_{VS_L}=\frac{3\pi}{2}, \\
A_{M}(\phi_{e'}^{T-cons}) &\leq & 0.062\, sin(\phi-\phi_{e'}) \,\, \mbox{$for$} \, \,\alpha_{VS_R}=0, \alpha_{VS_L}=0. 
 \eeq
 We see that the possible effect of up-down azimuthal asymmetry for the Majorana $\nu_{e}$'s is larger than in  the Dirac case. Moreover, there is a different dependence on the angle $\phi_{e'}$ in the case of the TRV and TRC, similarly  as for the Dirac neutrinos, so the precise measurement of maximal azimuthal asymmetry would  answer the question of whether the TRV takes place. 

\section{Conclusions}
\label{sec4}
We have shown that there is the distinction between the Dirac and Majorana  $\nu$'s  in the limit of vanishing $\nu$ mass, when the incoming $\nu$ beam is the mixture of LCh and RCh states, and has  the fixed direction of transversal component of the (anty)neutrino spin polarization with respect to the production plane. If the $\nu$ beam comes  from the polarized source ( e. g. PMDaR), where the exotic S, T interactions produce the $\nu$'s   with the right-handed chirality, while the V-A interaction generates the LCh $\nu$'s, the (anty)neutrino polarization vector may acquire the transversal component (both T-even and T-odd), which is left  invariant under Lorentz boost. Next, this left-right chiral mixture is scattered off the unpolarized electrons in the presence of both standard and exotic interactions.  The precise measurement of the azimuthal asymmetry of recoil electrons, generated by the interference terms between the standard  and  exotic
couplings, proportional to $\mbox{\boldmath $\eta_{\nu}^{ \perp}$}$,  would allow to distinguish between the Dirac and Majorana  $\nu$'s, and test the TRV. 
For the Majorana $\nu$'s, the upper limit on the expected magnitude of up-down azimuthal asymmetry is larger than for the Dirac case.   According to the SM, the angular distribution of recoil electrons  should be azimuthally symmetric in the massless  $\nu$ limit, and then  there is no difference between the Dirac and Majorana $\nu$'s. 
The basic difference between the both cases follows from  the absence of interference terms between  the standard and exotic tensor  interactions in the differential cross section for the Majorana $\nu$'s. 
The additional distinction  arises from the occurrence of  interference between the standard  and S couplings of the LCh  Majorana $\nu$'s. This type of interference annihilates  for the Dirac $\nu$'s. \\
It is also  important to note that the eventual effects connected with the neutrino mass and mixing  for the tests with a near detector are   inessential, see e.g. \cite{Sobkow}.  \\
It is relevant to stress that the azimuthal asymmetry measurements 
require    the very 
intense polarized  (anti)neutrino sources and large unpolarized (polarized) target of electrons (or nucleons),  and also long duration of experiment. 
To make the above tests  feasible,  the low-threshold, real-time detectors   should measure both  the polar angle and azimuthal angle of  outgoing electron momentum with a high resolution. It is necessary to point out that there is a real interest in the development of  low-threshold technology in the context of dark matter searches and the study of neutrino interactions.  The silicon cryogenic detectors,  the high purity germanium detectors 
\cite{nmm}, the semiconductor detectors \cite{semi} and the bolometers \cite{bolo} are worth mentioning. The two experiments aiming at the measurement of  
recoil electron scattering angle and of azimuthal angle, i. e. Hellaz \cite{Hellaz} and  Heron \cite{Heron},  have  also been proposed. Recently,  the interesting proposal for particle detection based on the infrared quantum counter concept  has emerged \cite{carugno}. 
Our studies are reported in hope that it may encourage the neutrino collaborations (e.g. KARMEN, PSI, TRIUMF,
BooNE, Borexino, Super-Kamiokande) working with the polarized muon decay, other  artificial polarized $\nu$ sources and neutrino beams to realize the measurements of the azimuthal asymmetry of recoil electrons. It seems to be a real  challenge, but new tests using the neutrino polarimeters  could shed much more light on  the $\nu$  nature,  detect the existence of the exotic couplings of interacting RCh  $\nu$'s and  the  non-standard  phases of TRV  than the present measurements based on  the electron (positron) observables and  energy spectrum of $\nu$'s.  
 Finally, it is worthy of  pointing out the fact that  in the massless $\nu$ limit, there are physical and in principle observable effects, coming  from the mixture between the LCh and RCh $\nu$ states (left- and right-handed helicity components) in the spin $1/2$ quantum state,  when  the exotic  S, T interactions coupling these two types of states exist  in the weak processes of $\nu$ production and detection.


\begin{thebibliography}{99}


\bibitem{SM} S. L. Glashow,  Nucl. Phys.  {\bf 22},  579 (1961);
 S. Weinberg,  Phys. Rev. Lett.  {\bf 19},    1264 (1967);
 A. Salam, in: N. Svartholm (Ed.),
 Elementary Particle Theory, Almquist and Wiksells, Stockholm,
1969;
  R. P. Feynman, M. Gell-Mann, Phys. Rev. {\bf 109},    193 (1958);
  E. C. G. Sudarshan, R. E. Marshak, Phys. Rev.  {\bf 109},   1860 (1958).
  \bibitem{Kayser} B. Kayser, R. E. Shrock, Phys. Lett. B {\bf  112}, 137 (1982). 
  \bibitem{Langacker} P. Langacker, D. London,  Phys. Rev. D {\bf  39}, 266 (1989). 
  \bibitem{Semikoz} V. B. Semikoz, Nucl. Phys. B {\bf 498},  39 (1997). 
  \bibitem{Pastor} S. Pastor, J. Segura, V. B. Semikoz, J. W. F.
Valle, Phys. Rev.  D {\bf 59} , 013004 (1998).
\bibitem{Barranco}  J. Barranco et al., Phys. Lett. B {\bf 739}, 343 2014. 
  \bibitem{Singh} D. Singh, N. Mobed and G. Papini, Phys. Rev. Lett. {\bf 97}, 041101 (2006). 
  \bibitem{Gutierrez} T. D. Gutierrez, Phys. Rev. Lett. {\bf 96}, 121802 (2006).
  \bibitem{Majorana} M. Doi et al., Phys. Lett. B {\bf 103}, 219 (1981); 
    W. C. Haxton et al., Phys. Rev. Lett. {\bf 47}, 153 (1981); 
    H. Ejiri, J. Phys. Soc. Jpn. {\bf 74}, 2101 (2005). 
         \bibitem{Rosen}S. P. Rosen, Phys. Rev. Lett. {\bf 48}, 852 (1982).
  \bibitem{Dass} G. V. Dass, Phys. Rev. D {\bf  32}, 1239 (1985).
  \bibitem{ndbd} D. Bogdan, A. Faessler and A. Petrovici, Neutrino masses and right-handed current in the neutrinoless double beta decay 76Ge  76Se + 2e*1, Volume 150, Issues 1-3, 3 January 1985, Pages 29-34.
  \bibitem{barion} A. Riotto and M. Trodden, Annu. Rev. Nucl. Part. Sci. {\bf 49},  35 (1999).
  \bibitem{Kobayashi} M. Kobayashi and T. Maskawa, Prog. Theor. Phys. {\bf 49},  652 (1973).
   \bibitem{Mulan} D. M. Webber et al.,  Phys. Rev. Lett. {\bf 106}, 041803 (2011).
  \bibitem{Michel} L. Michel and A. S. Wightman,   Phys. Rev. {\bf 98},  1190 (1955).
  \bibitem{Data} K.A. Olive et al. (Particle Data Group), Chin. Phys. C {\bf 38}, 090001 (2014).  
  \bibitem{Fetscher} W. Fetscher, Phys. Rev. D {\bf 49},  5945 (1994).
  \bibitem{Sobkow} W. Sobk\'ow, S. Ciechanowicz and M. Misiaszek, Phys. Lett. B {\bf  713}, 258 (2012). 
\bibitem{nmm} B.S. Neganov et al., hep-ex/0105083. 
\bibitem{semi} C. E. Aalseth et al., Phys. Rev. Lett. {\bf 106}, 131301 (2011).
\bibitem{bolo} C. Enss, Cryogenic Particle Detection (Springer-Verlag Berlin
Heidelberg, 2005).
  \bibitem{Hellaz} F. Arzarello et al.: Report No. CERN-LAA/94-19, College de France LPC/94-28,
1994. 
J. Seguinot et al.: Report No. LPC 95 08, College de France, Laboratoire de
Physique Corpusculaire, 1995. 
\bibitem{Heron} R. E. Lanou et al.: The Heron project, Abstracts of Papers of the American Chemical
Society 2(217), 021-NUCL 1999.
\bibitem{carugno} A. F. Borghesani et al., arXiv: 1506.07987 [physics.ins-det]. 
 
\end{thebibliography}
\end{document}